\documentclass{article}

\usepackage{graphicx}
\usepackage[round]{natbib}
\usepackage{epsfig}

\title
{
Improving Uncertain Climate Forecasts Using a New Minimum Mean Square Error Estimator for the Mean of the Normal Distribution
}

\author{Stephen Jewson and Ed Hawkins\\}

\begin{document}

\maketitle

\begin{abstract}
When climate forecasts are highly uncertain, the optimal mean squared error strategy is to ignore them.
When climate forecasts are highly certain, the optimal mean squared error strategy is to use them as is.
In between these two extremes there are climate forecasts with an intermediate level of uncertainty
for which the optimal mean squared error strategy is to make a compromise forecast.
We present two new methods for making such compromise forecasts, and show, using simulations,
that they improve on previously published methods.
\end{abstract}

\section{Introduction}
Forecasts for changes in climate vary greatly in terms of the ratio of the predicted signal to the estimated uncertainty around that signal.
For instance, numerical model derived predictions of changes in global temperature show a large ratio of signal to uncertainty, while forecasts of changes in local rainfall from the same models show a much smaller ratio of signal to uncertainty.
Forecasts with such different levels of uncertainty should be used very differently.
There are many ways one might model this mathematically, but one of the simplest is to consider the goal of the forecast
to be minimizing the estimated mean squared error (MSE) of the final prediction.
Such a framework leads one to ignore forecasts with a low value for the ratio between forecast signal and forecast uncertainty,
and to use forecasts with a high value for this ratio.
Interestingly, there is then a grey area in between, where a compromise forecast can be produced that achieves a lower mean squared error than either ignoring the forecast or using it in full. These compromise forecasts make better use of the information in climate model output since they
use this information to improve the forecast in parameter ranges where normally the model output would have to be ignored because of the high level
of uncertainty.
The statistical estimators used to make such compromise forecasts are sometimes known as minimum mean squared error estimators.

In~\citet{jewsonh09b}, we derived a simple method for making such a compromise forecast, and applied it to UK precipitation.
We called the method a `damped' forecast.
In this article we present two new methods for making damped forecasts based on new minimum mean square error estimators for the mean of the normal distribution.
We show, using simulations, that these new methods both outperform the simple damping method from~\citet{jewsonh09b}
within the most important parameter range.

In section~\ref{s2} we describe the mathematical set up we will use, and review the mean squared error performance of ignoring or using a climate forecast.
In section~\ref{s3} we then describe the simple estimator used in~\citet{jewsonh09b}, and its theoretical performance.
In section~\ref{s4} we use numerical methods to estimate the \emph{actual} performance of the mean squared error estimator from~\citet{jewsonh09b}.
In section~\ref{s5} we describe two new estimators, and compare the performance of all five methods discussed.
In section~\ref{s6} we summarise our results and discuss future directions.

\section{MSE performance of using or ignoring an uncertain climate forecast}
\label{s2}

Our mathematical setup follows that of~\citet{jewsonh09b}.
We consider values of a future climate variable to consist of a contribution from current climate, a contribution due to a change in climate, and noise. We write this as:
\begin{equation}
y=c+d+e
\end{equation}
We will assume, without loss of generality, that $c=0$, and so:
\begin{equation}
y=d+e
\end{equation}
Rather than deal with actual observed climate (which includes the noise) we simplify the algebra by considering future \emph{mean} climate only,
which we write as $\mu$.
Then we have:
\begin{equation}
\mu=d
\end{equation}

We imagine taking an estimate of the change in mean climate from a climate model (or climate model ensemble).
We write the estimated change as $\hat{d}$, and we assume that this is
unbiased ($E(\hat{d})=d$). We make an estimate of the current climate $\hat{c}$, and we assume that that is also unbiased.

An unbiased prediction of future mean climate is then given by:
\begin{equation}
\hat{\mu}=\hat{c}+\hat{d}
\end{equation}

We then assume that $\hat{c}=0$. This assumption does lose some generality, in that terms in the MSE of the estimate $\hat{c}$
drop out of the subsequent expressions for total MSE.
Essentially we are assuming that current climate can be estimated perfectly, which is clearly not true.
However, since we are only interested, in this article,
in the performance of predictions of the \emph{change} in climate, this assumption does not change any of our results
vis-a-vis the relative performance of different methods for predicting that change (and it simplifies the algebra a bit).

We then have:
\begin{equation}
\label{eq5}
\hat{\mu}=\hat{d}
\end{equation}

The mean squared error for this prediction of future mean climate is given by
\begin{equation}
\mbox{MSE}_1=E(\hat{\mu}-\mu)^2=E(\hat{d}-d)^2=E(\hat{d}-E(\hat{d}))^2=V
\end{equation}
where $V=V(\hat{d})$ is the variance of the estimate of the change in climate.
We don't discuss here how $V$ might be determined, but it could, for instance,
be derived from the spread of a climate model ensemble using the method we describe in~\citet{jewsonh09a}.

If instead of using climate model output in this way we decide to ignore it then our prediction is:
\begin{equation}
\hat{\mu}=0
\end{equation}

The mean squared error for \emph{this} prediction is:
\begin{equation}
\mbox{MSE}_2=E(\hat{\mu}-\mu)^2=E(0-d)^2=d^2
\end{equation}

We now consider the mean squared error normalised by $V$, which we write as NMSE (and NRMSE for the RMSE version).
We also define the ratio $r^2=\frac{d^2}{V}$.
For the two forecasts considered so far the NRMSE is:
\begin{equation}
\mbox{NRMSE}_1=1 \hspace{1in} \mbox{(using the climate model output)}
\end{equation}
and
\begin{equation}
\mbox{NRMSE}_2=\sqrt{\frac{d^2}{V}}=r \hspace{0.7in} \mbox{(ignoring the climate model output)}
\end{equation}

\begin{figure}[!ht]\begin{center}
\scalebox{0.8}{\includegraphics{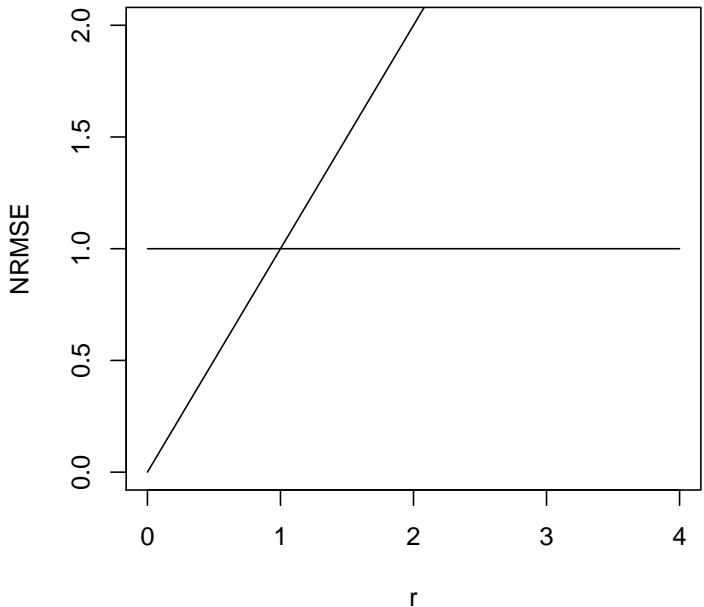}}
\end{center}\caption{
The normalised RMSE performance of
a) using a climate forecast (horizontal black line) and
b) ignoring a climate forecast (diagonal black line),
versus the ratio of signal to uncertainty.
}
\label{fig1}\end{figure}

Figure~\ref{fig1} shows the variation of NRMSE for these two forecasts versus $r$.
We see that for large values of $r$ ($r>1$) the option to use the forecast has the lower mean squared error,
while for
small values of $r$ ($r<1$) the choice of ignoring the forecast has the lower mean squared error.
This implies that for $r<1$ the climate model output should be ignored.
Among other benefits, the damping methods described below give a way of using climate model output to improve
forecasts even when $r<1$.

\section{A Simple Damped Estimator}
\label{s3}

In~\citet{jewsonh09b} we consider adjusting the prediction given in equation~\ref{eq5} using a \emph{damped} prediction:
\begin{equation}
\hat{\mu}=k \hat{d}
\end{equation}
for $k$ between zero and one.

We then derive the MSE optimal value of $k$, which is:
\begin{equation}
\label{eq1}
k=\frac{d^2}{d^2+V}
\end{equation}

The MSE for the prediction given by this optimal value of $k$ is

\begin{equation}
\mbox{MSE}_3=E(\hat{\mu}-\mu)^2=E(k\hat{d}-d)^2=\frac{d^2V}{d^2+V}
\end{equation}

(we've skipped a lot of details of these derivations, since the details are given in~\citet{jewsonh09b}).

The normalised RMSE is then given by:
\begin{equation}
\mbox{NRMSE}_3=\sqrt{\frac{d^2}{d^2+V}}=\sqrt{\frac{r}{1+r}}
\end{equation}

\begin{figure}[!ht]\begin{center}
\scalebox{0.8}{\includegraphics{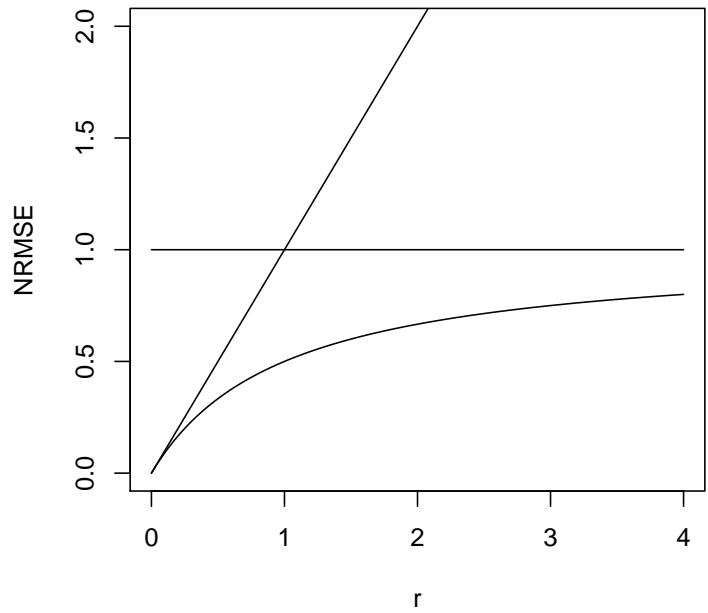}}
\end{center}\caption{
As figure 1, but now including the RMSE performance
of the ideal damping model (red line).
}
\label{fig2}\end{figure}

This is shown, along with $\mbox{NRMSE}_1$, and $\mbox{NRMSE}_2$, in figure~\ref{fig2}.

We see that \emph{in theory} the damped prediction is always better than the simpler predictions based on
either ignoring the model output or using it in full.
For very small values of $r$ it has more or less the same performance as ignoring the forecast,
for very large values of $r$ it has more or less the same performance as using the forecast,
and for intermediate values of $r \approx 1$ is performs markedly better than either.

However, there is a catch: the optimal value of $k$ is unknown, since the expression for $k$ given by equation~\ref{eq1}
depends on $d$ and $V$, which are both unknown.
For estimated values of $k$ the mean squared error performance is likely to be significantly worse than the theoretical ideal.
The performance of real estimates of $k$ can be determined using numerical methods, and that is described in the next section.

\section{MSE performance of the simple damped forecast}
\label{s4}

In~\citet{jewsonh09b} we used a plug-in estimator for $k$.
That is:
\begin{equation}
\hat{k}=\frac{\hat{d}^2}{\hat{d}^2+V}
\end{equation}

We now use numerical methods to determine how well this plug-in estimator performs, for a sample size of $n=10$.
Our methods work as follows:
\begin{itemize}
  \item For values of $r$ between 0 and 4 (with a step of 0.01), we simulate 1000 samples, each of size 10
  (each sample can be considered as an ensemble of climate model outputs)
  \item For each sample, we apply the three prediction methods described so far (ignore the sample, use the sample in full, or use the damped prediction)
  \item For each value of $r$ we calculate the NRMSE for the resulting predictions
\end{itemize}

\begin{figure}[!ht]\begin{center}
\scalebox{0.8}{\includegraphics{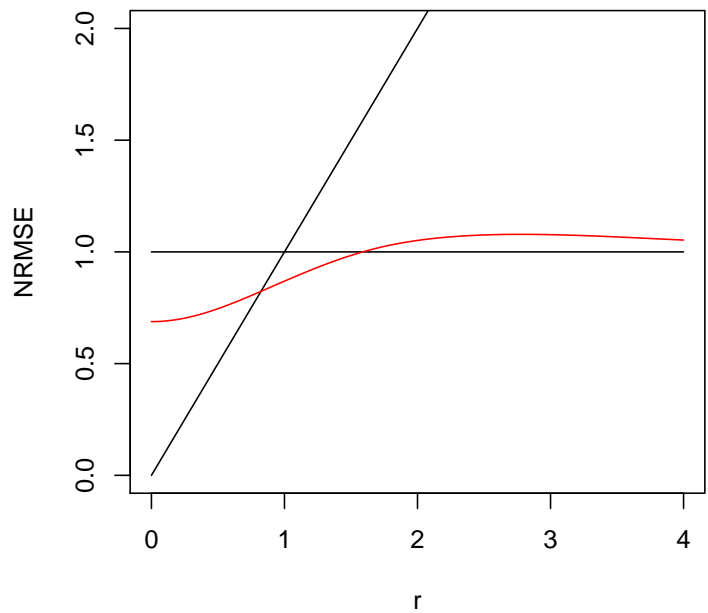}}
\end{center}\caption{
As figure 1, but now including the RMSE performance of
the plug-in estimator for the damping model (red line).
}
\label{fig3}\end{figure}

The results are shown in figure~\ref{fig3}.
We see that:
\begin{itemize}
  \item For very small values of $r$ the damped prediction performs less well than ignoring the forecast, but better than
  using the forecast in full
  \item For large values of $r$ the damped prediction performs less well than using the forecast in full, but much better than ignoring the forecast
  \item For intermediate values of $r$, around $r \approx 1$, the damped prediction performs better than either of the simpler forecasts.
\end{itemize}

We see that use of this damped forecast is not a panacea: it does not dominate the simpler methods for all values of $r$.
However, for certain values of $r$ it does improve the forecast. If one were forced to use just one of the three methods discussed
one would probably choose the damped forecast since it never performs too badly, unlike ignoring the forecast which performs
badly when the real signal is large, and unlike using the model output in full, which performs badly when the real signal is small.

\section{Two new minimum mean square estimators}
\label{s5}

We now move to the main topic of this paper, which is to present two new minimum mean square estimators.
The new estimators are based on the observation that the expression:
\begin{equation}
k=\frac{d^2}{d^2+V}
\end{equation}
is non-linear in $d$ and $V$, and so using a distribution of values for $d$ and $V$ may do better than using simple plug-in estimators
for each.

Motivated by Bayesian statistics, we therefore propose the following ad-hoc Bayesian estimator for $k$:
\begin{equation}
\hat{k}=\int \int \left( \frac{d^2}{d^2+V} \right) p(d,V|x) dd dV
\end{equation}

where $p(d,V|x)$ is the posterior probability for $d$ and $V$.
In other words, we consider all possible pairs of values of $d$ and $V$, we calculate the damping for each pair, and we average all
the damping coefficients together using a weighted average with the posterior probability as weights.

Our second estimator goes a step further and models the prediction directly, without going through an intermediate $k$, as:
\begin{equation}
\hat{\mu}=\int \int \left( \frac{d^3}{d^2+V} \right) p(d,V|x) dd dV
\end{equation}

We extend our numerical tests to include the performance of these two new estimators, as follows:
\begin{itemize}
  \item To determine the posterior, we use the standard uninformative prior for the normal distribution (which is also the Jeffreys' Prior).
  \item We evaluate the integrals using numerical integration. We discretize each dimension into 101 equal steps, over a range between minus and plus four standard errors.
\end{itemize}

\begin{figure}[!ht]\begin{center}
\scalebox{0.8}{\includegraphics{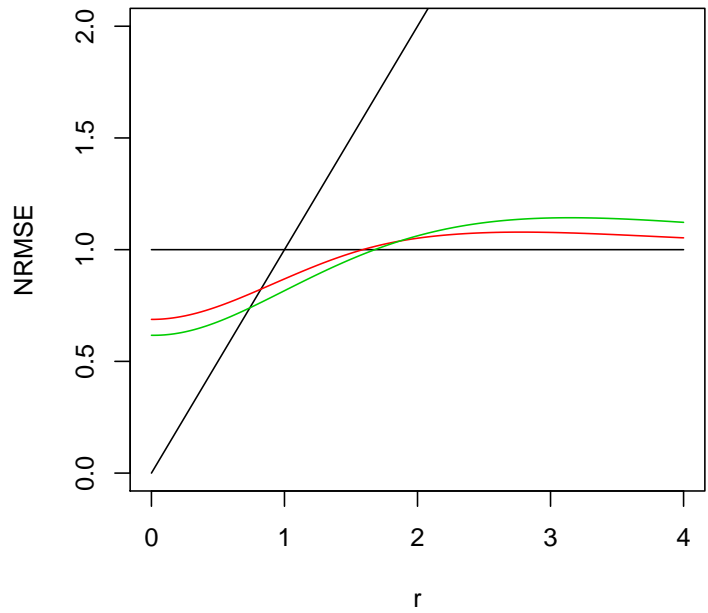}}
\end{center}\caption{
As figure 3, but now including the RMSE performance of
the first new estimator (green line).
}
\label{fig4}\end{figure}

Figure~\ref{fig4} shows the MSE performance of the first of these two estimators.
We see that it does better than the simple damping estimator up to just below $r=2$, but less well for larger values of $r$.
Over the range where the simple damping estimator beats the two simplest predictions, the new damping estimator clearly beats the simple
damping estimator.

\begin{figure}[!ht]\begin{center}
\scalebox{0.8}{\includegraphics{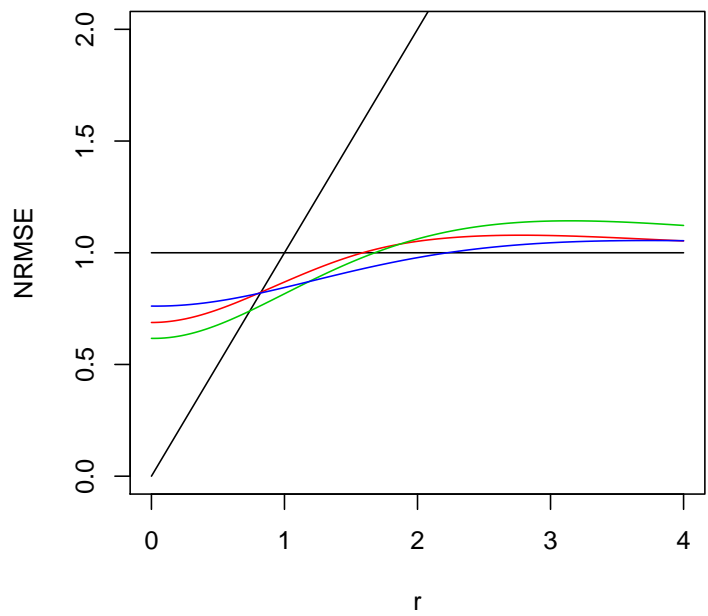}}
\end{center}\caption{
As figure 3, but now including the RMSE performance of
the second new estimator (blue line).
}
\label{fig5}\end{figure}

Figure~\ref{fig5} shows the MSE performance of the second of the two new estimators.
The second of the new estimators beats the simple damping estimator everywhere except for very small $r$.
Over the range where the simple damping estimator beats the two simplest predictions, the second of the new damping estimators clearly beats
the simple damping estimator.
However, the relationship between the two new estimators is complex. The second beats the first for larger values of $r$, and vice versa.

Overall, there are now four estimators which are the best, depending on the value of $r$:
\begin{itemize}
  \item For very small $r$, completely ignoring the forecast is best (for $r<0.74$).
  \item For slightly larger $r$, the first of the new damping estimators is best (for $0.73<r<1.19$).
  \item For $r$ larger again, the second of the new damping estimators is best (for $1.18<r<2.03$)
  \item And finally, for large $r$, using the forecast in full is best (for $r>2.02$).
\end{itemize}

The simple damping estimator is never the best, although it is not totally dominated by any of the other estimators.

\section{Summary}
\label{s6}

We have discussed the mean squared error performance of forecasts derived from the output of numerical climate models.
We have considered the mean squared error performance of 5 types of forecast derived from numerical climate model output:
\begin{itemize}
  \item Using the forecast as is
  \item Ignoring the forecast
  \item Damping the forecast using the damping scheme of~\citet{jewsonh09b}
  \item Damping the forecast using a new damping scheme
  \item Damping the forecast using a second new damping scheme
\end{itemize}

We have found that the new damping schemes outperform the original damping scheme of~\citet{jewsonh09b} over the most important range of
parameter values (which is the range over which the simple damping scheme beats ignoring the model output and using it in full).
Within this range the two new schemes each perform best for different ranges of the parameters.

The obvious next question is: how should one choose which of these methods to use in practice, given real climate model output?
Strictly speaking, this is \emph{not} answered by our results,
since we show the MSE versus an unknown parameter (the ratio of the unknown signal size to the unknown uncertainty around the signal).
One could decide which method to use by replacing the unknown parameter by an observed estimate, but this would be to ignore
uncertainty on that observed estimate. There may, therefore, be a better way to decide which method to use and when. We are looking into it.

There are also many other questions that arise from this work, such as:
\begin{itemize}
  \item Whether it would be worth considering metrics other than MSE
  \item Whether there are damping methods that work better than those proposed here
  \item Which climate forecasts, for what variables and at what lead times, fall into the various ranges of the parameter $r$.
\end{itemize}

%\bibliography{../../bib/jewson}
\bibliography{arxiv}

\end{document}